\newcommand{\halflengthimagesize}{0.49}
\newcommand{\shalflengthimagesize}{0.40}
\newcommand{\fulllengthimagesize}{0.98}

\documentclass[final,5p,times,twocolumn]{elsarticle}
\newcommand{\defaultimagesize}{\halflengthimagesize}
\newcommand{\sdefaultimagesize}{\shalflengthimagesize}

\usepackage{lineno,hyperref}
\usepackage{comment}
\usepackage{amsmath,amssymb}
\hypersetup{
    colorlinks=true,
    linkcolor=blue,
}
\usepackage[figurename=Fig., tablename=Table]{caption}
\usepackage{color}
\usepackage{soul}
\usepackage{cases}
\usepackage{ulem}
\usepackage{stfloats}
\usepackage{makeidx}
\usepackage{enumitem} 

\modulolinenumbers[1]
\journal{Journal of \LaTeX\ Templates}

\biboptions{sort&compress}

\begin{document}
\begin{frontmatter}

\title{First measurement of flux of the neutron background induced\\
by accelerated neutrinos at the J-PARC facility}

\newcommand{\AFFtus}{\address[AFFtus]{Department of Physics and Astronomy, Faculty of Science and Technology, Tokyo University of Science, Noda, Chiba 278-8510, Japan}}
\newcommand{\AFFynu}{\address[AFFynu]{Department of Physics, Yokohama National University, Yokohama, Kanagawa, 240-8501, Japan}}
\newcommand{\AFFokayama}{\address[AFFokayama]{Department of Physics, Okayama University, Okayama, 700-8530, Japan}}
\newcommand{\AFFkobe}{\address[AFFkobe]{Department of Physics, Graduate School of Science, Kobe University, Kobe, Hyogo 657-8501, Japan}}

\author[AFFkobe]{Hiroshi~Ito\corref{CorrespondingAuthour}}
\cortext[CorrespondingAuthour]{Corresponding author}
\ead{itoh.hiroshi@crystal.kobe-u.ac.jp}
\author[AFFynu]{Ryo~Shibayama}
\author[AFFtus]{Chisako~Ise}
\author[AFFynu]{Akihiro~Minamino}
\author[AFFokayama]{Yusuke~Koshio}
\author[AFFtus]{and Masaki~Ishitsuka}

\AFFkobe
\AFFynu
\AFFtus
\AFFokayama

\begin{abstract}
This article reports the measurement of the neutron background flux accidentally induced by the neutrino beam at the J-PARC facility in Japan.
In particular, neutrino-nucleus neutral-current quasi-elastic (NCQE) scattering events, $\nu~(\bar{\nu})+\rm{N}\to\nu~(\bar{\nu})+\rm{N'}$, can be obscured by a massive background of neutron-nucleus scattering events.
The neutrons are produced within the materials (such as sand and concrete) located between the beam dump and the experimental area where the detectors are placed.
We measured the accidental neutron events at the J-PARC neutrino facility using BGO and liquid scintillation detectors, with a plastic scintillation detector serving as an active veto counter.
Based on a neutrino-mode data set of \(2.972 \times 10^{20}\) POT, we observed 88 neutron-induced recoil proton events within an electron-equivalent recoil energy range of 0.98--11.60~$\rm MeV_{ee}$, selected via pulse-shape discrimination in the liquid scintillation detector.
Accounting for the detection efficiency and resolution, and through comparison with simulations, the neutron flux was determined to be\([1.45^{+0.22}_{-0.24}~\text{(stat.)} \pm 0.55~\text{(sys.)}] \times 10^{-7}~\text{cm}^{-2}\,\text{s}^{-1}\,\text{POT}^{-1}\),assuming an exponential neutron energy spectrum. This result will contribute to the evaluation of neutron backgrounds for neutrino experiments at the J-PARC facility.
\end{abstract}

\begin{keyword}
Neutrino experiment\sep
Neutron\sep
BGO detector\sep 
Liquid Scintillation detector
\end{keyword}
\end{frontmatter}


%
%
\section{Introduction}
\label{Introduction}

Neutrino-nucleus scattering events are of great interest; they serve as a powerful probe into the fine structure of nuclei and provide unique insights distinct from those obtained through other interactions.
In particular, the neutrino-nucleus neutral-current quasi-elastic (NCQE) interaction is mediated by the $Z$ boson, which leaves the charges of the leptons in the initial and final states unchanged. This interaction is expressed as follows:$\nu_l (\bar{\nu}_l)+ N \to \nu_l (\bar{\nu}_l)+ N'$, where $N$ and $N^{\prime }$ denote a nucleon (neutron or proton) bound in the nucleus, and $l$ represents the lepton flavor ($e$, $\mu$, or $\tau$).
In the NCQE interaction, 
we can measure proton and neutron recoils.

The NCQE cross section is predicted to depend on the form factor as a function of the axial strange coupling constant, $g_{\mathrm{A}}^{s}$. Specifically, varying $g_{\mathrm{A}}^{s}$ from 0 to $-0.3$ is predicted to enhance the cross section of the proton channel ($\nu(\bar{\nu})+p\to\nu(\bar{\nu})+p$) while suppressing that of the neutron channel ($\nu(\bar{\nu})+n\to\nu(\bar{\nu})+n$), while the total cross section (the sum of the proton and neutron channels) remains invariant.

The BNL E734 experiment measured the cross section of the \(\nu(\bar{\nu})\)-hydrocarbon NCQE interaction, determining the coupling constant to be \(g_{\rm A}^s = -0.15 \pm 0.07\)~\cite{PhysRevC.48.761}. The MiniBooNE collaboration also reported the cross-section measurement for the \(\nu(\bar{\nu})\)-hydrocarbon NCQE interaction; subsequently, using the MiniBooNE data, Golan et al. extracted a value of \(g_{\rm A}^s = -0.4^{+0.5}_{-0.3}\)~\cite{PhysRevC.88.024612}. Furthermore, the KamLAND experiment constrained the \(g_{\mathrm{A}}^{s}\) factor to be \(-0.14_{-0.26}^{+0.25}\) using atmospheric \(\nu \)-hydrocarbon NCQE interaction events~\cite{PhysRevD.107.072006}.

On the other hand, $\nu$-oxygen NCQE is one of the main backgrounds for the diffuse supernova neutrino background (DSNB) search. In the recent search results reported by Super-Kamiokande (SK) using the gadolinium-loaded phase, the remaining events were consistent with the estimated background~\cite{SK2023-SRN}. The NCQE background events in SK are estimated using the following models: the HKKM model for the atmospheric neutrino flux~\cite{HKKM2011, HKKM2007}, NEUT 5.4.0.1 for the neutrino interaction~\cite{NEUT2009, NEUT2021}, and the Ankowski et al. model for the nuclear de-excitation $\gamma$ rays from the NCQE interaction below an energy of 16 MeV~\cite{Ankowski}. The total systematic uncertainty in estimating the number of NCQE events is estimated to be 68\% below 15.49 MeV and 82\% above 15.49 MeV.

The T2K collaboration measured the \(\nu \)-\({}^{16}\)O NCQE cross section to be \(1.70\times 10^{-38}~{\rm cm^2 /oxygen}\) with a statistical uncertainty of \(\pm10\%\) and a systematic uncertainty of \({}_{-22\%}^{+30\%}\) in the neutrino mode~\cite{PhysRevD.100.112009}. For the antineutrino mode, the cross section was determined as \(0.98\times 10^{-38}~{\rm cm^2 /oxygen}\) with \(\pm16\%~{\rm (stat.)}\) and \(^{+27\%}_{-19\%}~{\rm (syst.)}\)~\cite{PhysRevD.100.112009}. This result was obtained using SK as the far detector. To reduce the NCQE model uncertainty, larger statistics are required. On a short baseline, high-statistics NCQE measurements can be achieved by using relatively compact detectors. The main background for the NCQE proton channel is expected to be neutrino-induced neutron-proton elastic scattering events.


In this paper, we first briefly outline the test experiment in Section~\ref{sec.T101}, followed by a description of the equipment, facility, setup, and simulation in Section~\ref{sec.MateMetho}. We then detail the analysis method in Section~\ref{sec.analysis} and discuss the experimental results in Section~\ref{sec.result_discusstion}. Finally, we conclude the paper in Section~\ref{sec.conclusion}.

\section{J-PARC T101 test experiment}
\label{sec.T101}

A new short-baseline physics experiment to study NCQE interactions is being planned based on the detection of nuclear recoil events. This project is inspired by recent measurements of the quenching factor for oxygen nuclear recoils in BGO crystals~\cite{Ommura_2023}, which observed neutron-oxygen elastic scattering events. However, the primary background is anticipated to be neutron-nucleus inelastic scattering events.

Therefore, we aim to evaluate the accidental neutron background in the BGO crystal by determining the flux of neutrons induced by accelerator-produced neutrinos at the J-PARC neutrino facility~\cite{T101_proposal}.

Under the J-PARC T101 test experiment, we conducted measurements over three separate periods using different experimental setups. The exact setup, protons on target (POT), and key characteristics of each setup are summarized in Table~\ref{table.period}.

Setups 1 and 2 correspond to the distinct beam operation periods in February and June, respectively. During the maintenance period in June, the configuration was updated to Setup 3 by incorporating additional BGO detectors to accommodate potential measurements of neutrino charged-current scattering events.

   %
    %
    \begin{table*}[htbp]
    \centering
    \caption{J-PARC T101 dataset}
    \begin{tabular}{c c c c c}
    \hline
    \hline
      & Period of physics run& beam power & P. O. T. & Characteristics of the setup\\    
    \hline
    \hline
    Setup-1 & 12--23, Feb. 2024 & 650~kW & $9.23\times10^{19}$ & Detectors on the table\\
    
    Setup-2 & 6--19, Jun. 2024 & 710--800~kW & $13.70\times10^{19}$ & Add aluminum frame to fix\\
    
    Setup-3 & 19--28, Jun. 2024 & 700--800~kW  & $6.79\times10^{19}$ & Add BGO crystal\\    
    \hline
    Total &&& $29.72\times10^{19}$ & \\
    \hline
    \hline
    \end{tabular}
    \label{table.period}
    \end{table*}

\section{Material and Methods}
\label{sec.MateMetho}

This section details the experimental configurations of the T101 experiment, including the equipment (detectors and front-end DAQ), the experimental site at the J-PARC facility, the setups, beamtime, accumulated datasets, and simulation.

\subsection{Equipment}
\label{sec.Equipment}

\subsubsection{Plastic scintillation detector}

A plastic scintillation detector (hereafter PlasSci) is used as an active veto counter for charged particles. The size of the plastic scintillator is \(200\times200\times10~{\rm mm}^3\). A photomultiplier tube (PMT; R329, Hamamatsu Photonics) is optically coupled to the scintillator via an acrylic light guide.

\subsubsection{BGO detectors}
\label{sec.BGO}

We used two BGO (\(\rm{Bi}_4\rm{Ge}_3\rm{O}_{12}\), density \(7.13~{\rm g/cm}^3\)) crystal detectors: a cylindrical crystal, 5.08~cm in diameter and 5.08~cm in length (hereafter BGO-1), and a cubic crystal with dimensions of \(30\times30\times30~{\rm mm}^3\) (hereafter BGO-2). Both crystals are optically coupled to PMTs on both sides and light-shielded using aluminized Mylar and a black sheet.

Although the BGO detectors are not used in the present neutron-flux analysis, BGO-1 (and, in setup~3, BGO-2) serves as the tagging detector for the energy calibration of the LiqSci detector, described in Section~\ref{sec.LiqSci} and detailed in Appendix~A. In setup~3, BGO-2 was additionally installed to search for neutrino-nucleus charged-current interactions via a triple coincidence of BGO-1, BGO-2, and LiqSci (Section~\ref{sec.setup}).

\subsubsection{Liquid scintillation detector}
\label{sec.LiqSci}

A liquid scintillation detector (hereafter LiqSci) is used for neutron identification based on pulse-shape discrimination (PSD) of the signal waveform. It consists of a cylindrical vessel (7.62~cm in diameter and 7.62~cm in length) filled with a liquid scintillator (BC-501A, Saint-Gobain) and a photomultiplier tube (9822B, ET Enterprises) optically coupled to the window of the vessel. The specifications of this detector are identical to those of the liquid scintillator detector described in Ref.~\cite{Ommura_2023}.

The energy calibration of the LiqSci detector was performed using \(^{22}\)Na and \(^{60}\)Co gamma-ray sources placed between the BGO and LiqSci detectors. By tagging photoelectric-absorption events (511, 1173, or 1332~keV) in the BGO detector, we selected events in which the counterpart gamma ray entered the LiqSci detector, enabling a direct, source-tagged comparison between the experimental data and a simplified simulation. The resulting Compton-edge spectra were fitted to determine the energy-to-charge conversion coefficient \(k\) and the energy resolution, parameterized as \(\sigma=\sigma_1 E^{1/2}+\sigma_0\), where \(E\) is the deposited energy in units of MeV electron equivalent. Because the high voltage supplied to the LiqSci detector differed between setups (\(-1300\)~V for setups~1 and 2, and \(-1329\)~V for setup~3), the calibration was performed separately for the two voltage settings; the resulting best-fit values of \(k\), \(\sigma_0\), and \(\sigma_1\) are summarized in Table~\ref{table.calib.LS}.

The neutron identification (PSD) measurements were performed at RCNP after the beamtime. Further details of the energy calibration and the PSD calibration, including the calibration setup and fit results, are given in Appendices~A and B, respectively. The evaluated energy resolution and PSD efficiency were incorporated into the detector response function within the simulation.

\subsubsection{Front-end electronics and data acquisition}

\begin{figure}[hbtp]
    \centering
    \includegraphics[width=\defaultimagesize\textwidth]{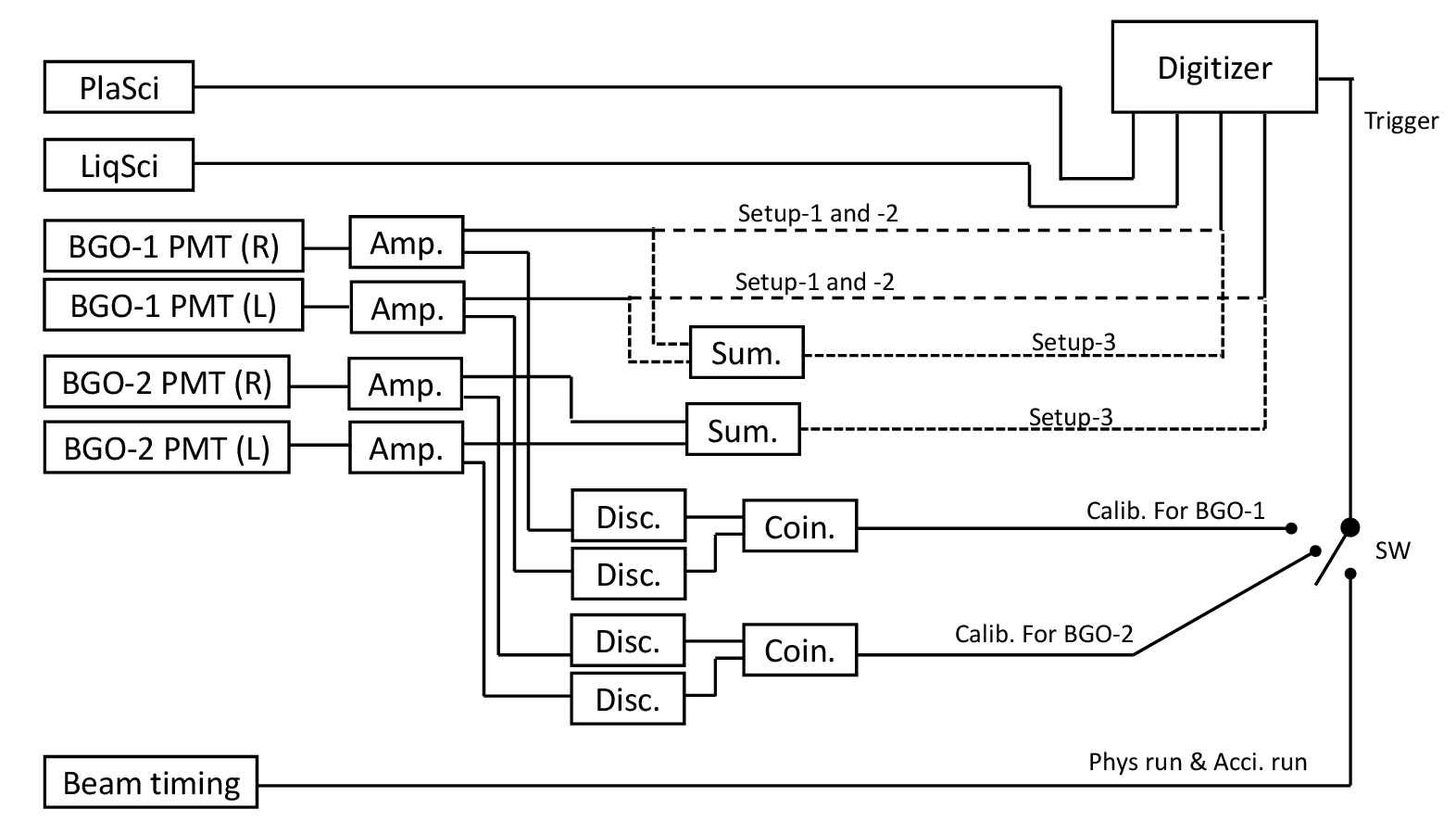}
    \caption{
    Front-end data acquisition system overview.
    In setup-1 and -2, dashed lines were used for BGO signals.
    In setup-3, dotted lines were used for BGO signals.
    External trigger mode has been switched at SW.
    }
    \label{fig.DAQ}
    \end{figure}

The front-end electronics are illustrated in Fig.~\ref{fig.DAQ}. The analog signal waveforms from PlasSci, the two BGO detectors, and LiqSci are recorded by a 4-channel, 12-bit, 250~MS/s digitizer (CAEN DT5720). The data acquisition (DAQ) window is set to \(100~\mu{\rm s}\) for the physics and accidental background runs, and \(12~\mu{\rm s}\) for the calibration run. For setups 1 and 2, the analog signals from both PMTs of BGO-1 are individually amplified and recorded. In setup~3, however, the analog signals from the two PMTs of BGO-1 (and BGO-2) are amplified and then summed into a single channel, which is then recorded.

    %
    %

During setups 1 and 2, the two PMTs coupled to BGO-1 were digitized independently to enable individual gain matching and energy calibration. In setup 3, their signals were summed prior to digitization in order to conserve DAQ channels for additional detectors, since stable PMT gain calibrations had already been established by that stage.

In the calibration run, a trigger signal was set both phototubes detected scintillation light in the BGO crystal for BGO-1 and -2.
The trigger system can be configured differently for the physics, accidental background, and calibration runs. In the physics run, the trigger signal for the digitizer was synchronized with the timing of the proton beam entering the neutrino beamline from the J-PARC Main Ring. The same trigger was also used for the accidental background run during beam-off periods. In the calibration run, the trigger was generated when both PMTs simultaneously detected scintillation light in the BGO-1 or BGO-2 crystals.

    %
    %
    \begin{figure*}[hbtp]
    \centering
    \includegraphics[width=\fulllengthimagesize\textwidth]
    {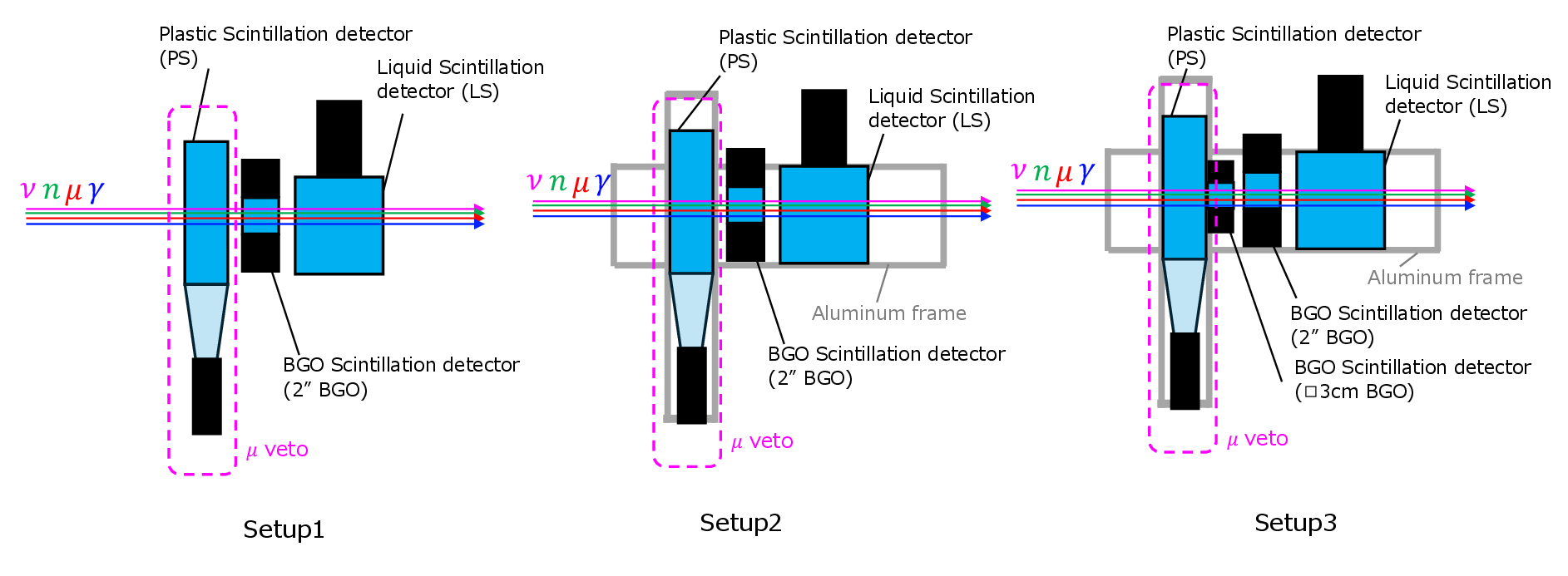}
    \caption{
    Experimental setups 1, 2, and 3.
    The plastic scintillation detector is commonly used to veto muon events.
    The aluminum frame is added from setup 1 to 2.
    The small-sized BGO detector is added from setup 2 to 3.
    }
    \label{fig.setup}
    \end{figure*}

    %
    %
    \begin{figure*}[hbtp]
    \centering
    \includegraphics[width=\fulllengthimagesize\textwidth]{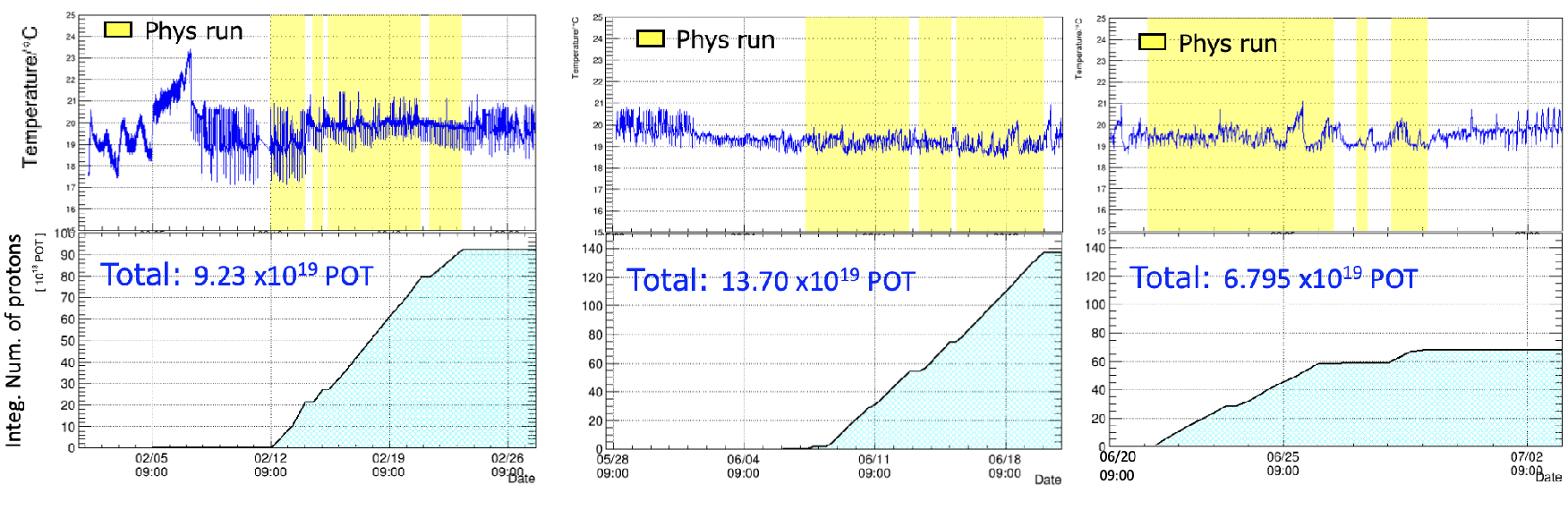}
    \caption{
    Monitored 
    room temperature in setup-1, -2, and -3 (top). Yellow bands show the beam on timing as physics runs for these setups.
    The amounts of data obtained are expressed as the integrated number of protons for setups 1, 2, and 3 (bottom). In setup-1, 2, and 3, total amounts are 9.23, 13.70, and 6.795 $\times10^{19}$ POT, respectively.
    }
    \label{fig.dataset}
    \end{figure*}
    
\subsection{J-PARC neutrino facility}
\label{sec.J-PARC}

\begin{figure}[hbtp]
    \centering
    \includegraphics[width=\defaultimagesize\textwidth]{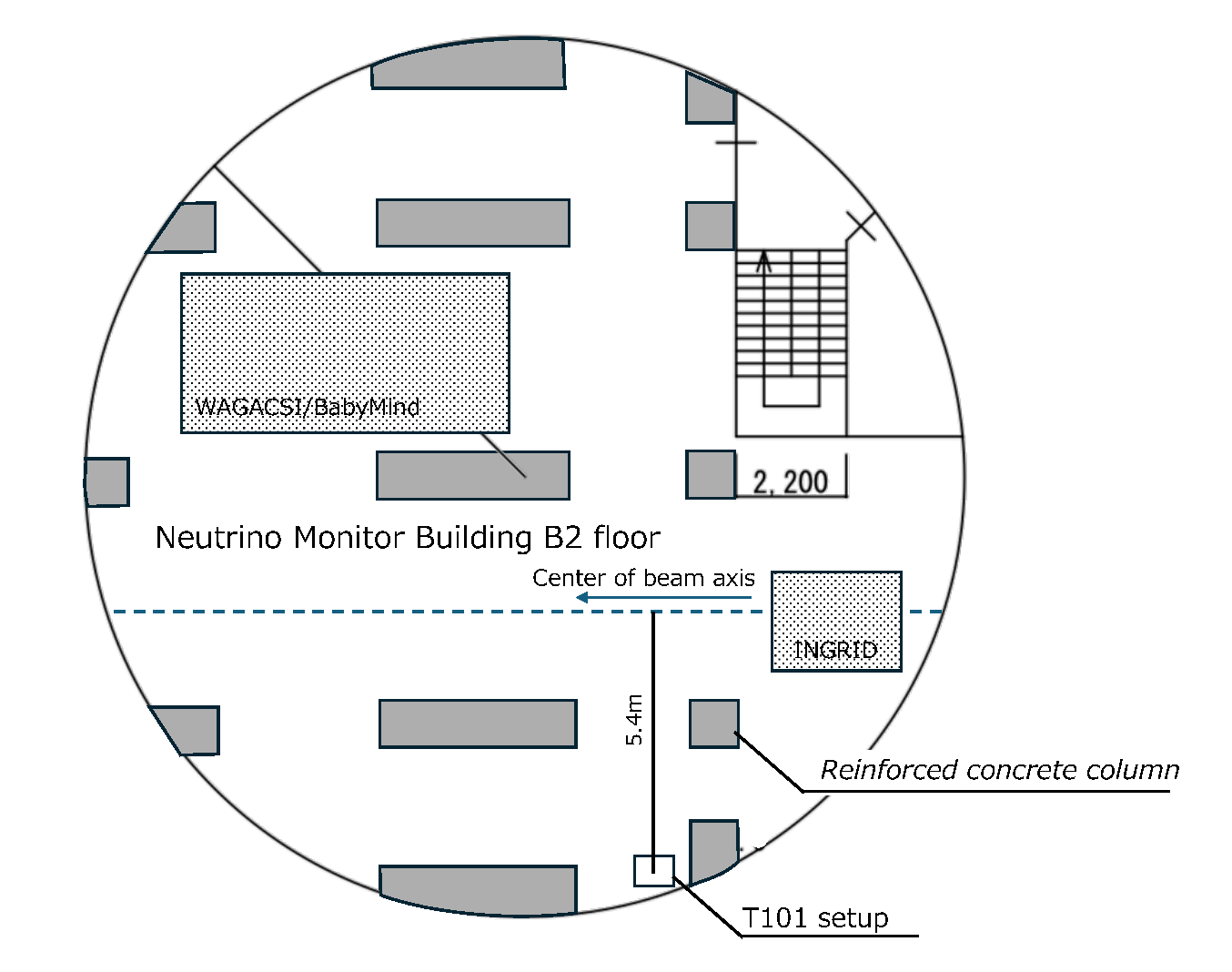}
    \caption{
    Schematic of the B2 floor of the Neutrino Monitor Building.
    }
    \label{fig.NMB2}
    \end{figure}

The J-PARC Main Ring accelerates a 30-GeV proton beam with a repetition cycle of 1.36~seconds, operating at a beam power of 600--800~kW. The neutrino beams are delivered per acceleration cycle in units called "spills." Each spill contains eight bunches within a duration of approximately \(5~\rm{\mu s}\). The neutrino beamline is composed of two sections: the primary and secondary beamlines (for further details, see Ref.~\cite{PhysRevD.87.012001}). The generated neutrino beam reaches the neutrino monitor (NM) building at J-PARC. In the underground experimental hall (B2 floor), the WAGASCI~\cite{WAGACSI2016}, NINJA~\cite{NINJA2017a, NINJA2017b}, and INGRID detectors are located at a \(1.5^{\circ }\) off-axis angle.
%
We performed our test experiment at a location 5.4~m south of the beam axis center on the B2 floor of the NM building, as shown in Fig.~\ref{fig.NMB2}.

    %
    %

\subsection{Setup}
\label{sec.setup}

We performed measurements using three different setup configurations across the setups, as shown in Fig.~\ref{fig.setup}. In setup 1, the PlasSci, BGO-1, and LiqSci detectors were aligned on a table along the neutrino beam axis. In setup 2, these detectors were secured within an aluminum frame, which slightly shifted their positions. In setup 3, BGO-2 was inserted between the PlasSci and BGO-1 detectors. This setup was designed to search for neutrino-nucleus charged-current interactions within the BGO crystals, which were identified via a triple coincidence of BGO-1, BGO-2, and LiqSci under a PlasSci veto condition. Since our digitizer features only four input channels, the two analog signals from the PMTs of BGO-1 (and BGO-2) were summed into a single channel.

Throughout all setups, we monitored the room temperature because the light yield of BGO crystals is strongly temperature-dependent (\(-1.69\%/^\circ{\rm C}\) within 10--30\(~^{\circ }\mathrm{C}\); see the top panel of Fig.~\ref{fig.dataset}). Temperature corrections were applied to the BGO energy values used in the LiqSci energy calibration (Appendix~A), whereas the light yield of the liquid scintillator itself exhibits negligible temperature dependence, having no impact on the results.

\subsection{Monte Carlo simulation}
\label{sec.MC}

We utilized the Geant4.11.1 Monte Carlo simulation package to model particle scattering within the detector. After the simulation data were tuned to match the format of the experimental data using the calibration results, the selected events were compared.

\subsubsection{Detector response}
\label{sec.MC.detector}

To determine the detector response, calibration tests were conducted using gamma-ray and neutron sources. The resulting energy resolution (detailed in Appendices A and B) was incorporated into the simulation to allow for a direct comparison with the observed energy.

\subsubsection{Quenching factor}
\label{sec.MC.QF}

The scintillation light yield for nuclear recoils tends to be smaller than that for electron recoils. The ratio of the nuclear recoil light yield to the electron recoil light yield is defined as the quenching factor.
For the LiqSci simulation, we adopted the quenching factor measurement results for carbon recoils~\cite{YOSHIDA2010574} and proton recoils~\cite{ConnorAwe2021} as a function of the recoil energy.

The quenching factor may depend on the specific scintillator composition. The KamLAND liquid scintillator (mixture of normal-dodecane, pseudocumene, and PPO) and plastic scintillators (EJ-260/EJ-228) have reported $kB$ factors of $(7.8\pm0.13)\times10^{-3}~{\rm g/cm^2/MeV}$ \cite{YOSHIDA2010574} and $(8.7\pm0.9)\times10^{-3}~{\rm g/cm^2/MeV}$ \cite{ConnorAwe2021}, respectively. 

These values are consistent within uncertainties with the recent experimental value of ${kB\sim9.3\times10^{-3}~{\rm g/cm^2/MeV}}$ reported for EJ-301 (equivalent to the xylene-based BC501A) \cite{NIMA1042_167405}. Because the QF curves from \cite{YOSHIDA2010574} and \cite{ConnorAwe2021} agree within error margins, and Ref.~\cite{ConnorAwe2021} carries a large measurement error at proton recoil energies, any uncertainty arising from the different scintillator cocktails is fully covered by this error. Consequently, rather than applying explicit corrections, this experimental error range from Ref.~\cite{ConnorAwe2021} was treated as the dominant contribution to the systematic uncertainties.

\subsubsection{Neutron energy spectrum}
\label{sec.MC.neutron_spectrum}

It is considered that 
neutrons measured in our detector
are induced by accelerator-produced neutrinos within the sand wall surrounding the experimental hall. Modeling the resulting neutron flux is challenging due to limited data on neutron production points, their energy deposition before escaping the sand, and the sand's precise composition and density. 
At J-PARC, the muon neutrino energy spectrum peaks at ${\sim0.7~\rm GeV}$ at a 1.5\({}^{\circ }\) off-axis angle,
and
secondary neutrons exhibit kinetic energies up to the sub-GeV scale.

The energy spectrum of neutrons incident on the detector is considered to be a superposition of energy distributions from neutrons generated throughout the sand.
Thus, we assumed the neutron energy spectrum to be an exponential function of the neutron kinetic energy \(E_{n}\),
\begin{equation}
{\it\Gamma}(E_n)=A \exp{\left(-\frac{E_n}{\lambda}\right)},\label{eq1}
\end{equation}
where \(A\) is the normalization amplitude and \(\lambda \) is the decay constant.
In this study, we first generated a dataset with neutrons uniformly distributed in initial kinetic energy between 1 and 10,000~MeV. We then re-weighted the spectrum shape by treating \(A\) and \(\lambda \) as free parameters in the fit to the experimental data.

\subsubsection{Neutron direction}
\label{sec.MC.neutron_direction}

The initial direction of the neutrons also cannot be modeled, as they can be produced downstream of the walls within the experimental hall, in which case they would travel upstream. Therefore, we assumed an isotropic distribution for the neutron directions.

\section{Analysis}
\label{sec.analysis}

The primary analysis strategy is to subtract the events captured during the beam-off time window from those in the beam-on time window to evaluate the net number of events. In this section, we detail the definitions of the beam-on and beam-off periods, the proton-recoil event selection in the LiqSci, the charged-particle event rejection using the PlasSci, and the re-weighting fit method utilized to determine the neutron flux.

\subsection{Beam timing}

\begin{figure}[hbtp]
    \centering    \includegraphics[width=\defaultimagesize\textwidth]{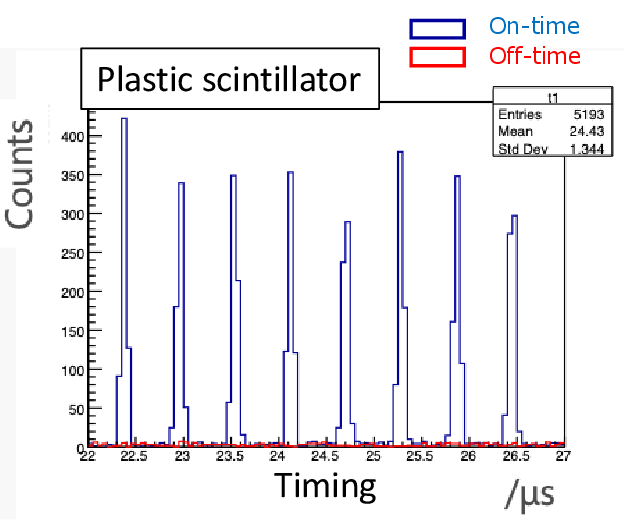}
    \caption{
    Hit time distribution in the plastic scintillation detector. Off-time events (red) are shifted to the on-time region and overlaid with the on-time events (blue).
    }
    \label{fig.beam_bunches_PS}
    \end{figure}

Accelerator-produced neutrinos interact with the sand material upstream of the experimental hall, producing muon backgrounds from the wall surface (referred to as "sand muons") that are synchronized with the beam timing. In the physics run, these sand-muons were observed as eight distinct beam bunches, as shown in Fig.~\ref{fig.beam_bunches_PS}. The on-time and off-time regions were defined as \(22\text{–}27~\mu{\rm s}\) and \(5\text{–}10~\mu{\rm s}\) within the recorded time window, respectively.

    %
    %

\subsection{Multiple hit pulse}

Within the \(100~\mu{\rm s}\) DAQ window, multiple pulses can be detected in the liquid scintillation detector. For the on-time region, we account for these multiple hits to accurately evaluate the event rate. The pulse separation time limit is estimated to be \(0.4~\mu{\rm s}\) with a threshold energy of 0.98 MeV electron-equivalent ($\rm {MeV}_{ee}$). The pulse multiplicity reaches up to 4, following an exponential distribution.


The pulse multiplicity follows a Poisson distribution. Owing to the small average number of events per DAQ window ($\lambda \ll 1$), the multiplicity distribution exhibits an exponential-like suppression as the multiplicity increases. In the current dataset, the observed pulse multiplicity reached up to 4, and no events with a multiplicity of 5 or higher were observed due to the limited statistics.

\subsection{Pulse-shape discrimination}
\label{sectoin.PSD}

The difference in the signal waveform shape of the LiqSci detector allows us to distinguish neutron events from gamma-ray events, a technique known as pulse-shape discrimination (PSD). In this analysis, the PSD method is fundamentally identical to that used in Refs.~\cite{Ommura_2023, ITO2023168701}. However, this specific approach is not applicable to proton recoils in the sub-GeV region. Therefore, to calibrate the PSD for this region, we conducted measurements by irradiating the LiqSci detector with a 65-MeV neutron beam at the RCNP facility at Osaka University. The details of this calibration are described in Appendix~B.

\subsection{Rejection of charged particle event}

The PlasSci also registers multiple hit events, where muons spanning two or three bunches within the same spill are observed. To reject charged-particle events, the PlasSci signals are digitized into a logic value of 1 or 0, using a threshold of 50 mV in pulse height and a time resolution of \(0.1~\mu{\rm s}\) within the \(20\text{–}30~\mu{\rm s}\) window.
This 50~mV threshold is sufficiently low compared to the minimum ionizing particle energy, ensuring a rejection efficiency of more than 99\%.
We required that this logic value be zero within the corresponding time frame for each LiqSci event, thereby rejecting the charged-particle background.

\subsection{Proton recoil event selection}

\begin{figure}[htp]
    \centering
    \includegraphics[width=\defaultimagesize\textwidth]   
    {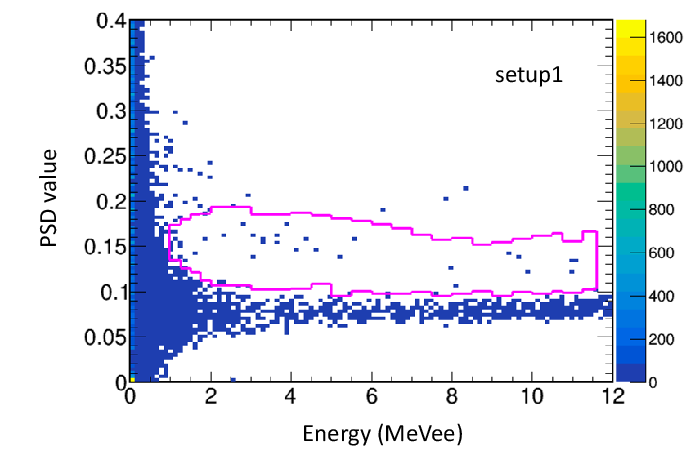}
    \caption{
    Distribution of deposited energy versus the PSD value for setup 1. Dots represent the experimental data, and the magenta closed line indicates the cut criteria boundary for proton-recoil events.
    }
    \label{fig.PSD.plot}
    \end{figure}

The distribution between the PSD values and deposited energy observed by the LiqSci in setup 1 is shown in Fig.~\ref{fig.PSD.plot}. Proton-recoil events were selected based on cut criteria determined by the calibration at the RCNP neutron facility (detailed in Appendix B). At energies below \(2~{\rm MeV_{ee}}\), gamma-ray events extend over a wide range of PSD values, overlapping with and thus restricting the selection region for proton-recoil events. On the other hand, when the PSD value is 0.2 or higher, recoiling charged particles heavier than protons were observed in the RCNP calibration. In this study, due to the limited literature on the quenching factors of heavy charged particles in liquid scintillators, we applied an upper limit on the PSD values to eliminate these heavy charged-particle events.

    %
    %
    
\subsection{Re-weighted spectrum fit}

To optimize the simulated energy deposition spectrum of neutrons in the LiqSci to match the experimental data, a reweighting procedure was applied based on the initial neutron energy at production, $E_n$. 

The reweighting procedure was carried out through the following steps:
\begin{enumerate}
    \item For each simulated MC event $i$, the initial neutron energy $E_{n,i}$ and the corresponding energy deposited in the LiqSci (after applying the detector response and quenching factors described in Section~\ref{sec.MC}) were recorded.
    \item A weight $w(E_{n,i}; A, \lambda)$ was assigned to each event as a function of its initial energy $E_{n,i}$, using the exponential parametrization:
    \begin{equation}
        w(E_{n,i}; A, \lambda) = A \exp(-\lambda E_{n,i}),
    \end{equation}
    where $A$ is the normalization amplitude and $\lambda$ is the decay constant.
    \item The re-weighted energy deposition spectrum in the LiqSci was reconstructed by filling the histograms of the deposited energy, with each event weighted by $w(E_{n,i}; A, \lambda)$.
    \item The optimal parameters $A$ and $\lambda$ were determined by minimizing the $\chi^2$ value between the re-weighted simulated energy deposition spectrum and the measured experimental spectrum.
\end{enumerate}

We applied this re-weighted spectrum fit to the spectra of three experimental setups, all of which involved proton recoil events in the LiqSci.
The \(\chi ^{2}\) value for the fit is defined as follows:
\begin{equation}
\chi^2 = \sum_i 2 \left[ N_i - M_i(A,\lambda) + N_i \ln\left(\frac{N_i}{M_i(A,\lambda)}\right) \right],
\end{equation}
 where \(N_{i}\) is the experimental data, \(M_i(A,\lambda)\) is the re-weighted simulation prediction, and \(i\) denotes the bin index of the spectrum.
The \(1\sigma\) uncertainties (68.3\% confidence level) for \(A\) and \(\lambda \) were determined using \(\Delta \chi^2=2.31\) for the two variables.

\section{Result and Discussion}
\label{sec.result_discusstion}

\subsection{Deposited energy spectrum in Liquid scintillator}
\label{sec.ene.spectrum}

We observed proton-recoil-like events in the energy range of 0.98--11.60~MeV. The net event rates were evaluated to be $(3.3\pm0.6)\times10^{-19}$, $(2.3\pm0.4)\times10^{-19}$, and \((2.5\pm0.7)\times10^{-19}~{\rm events~POT^{-1}}\) for setups 1, 2, and 3, respectively, which are consistent with each other.

\begin{figure}[htp]
    \centering
    \includegraphics[width=\sdefaultimagesize\textwidth]{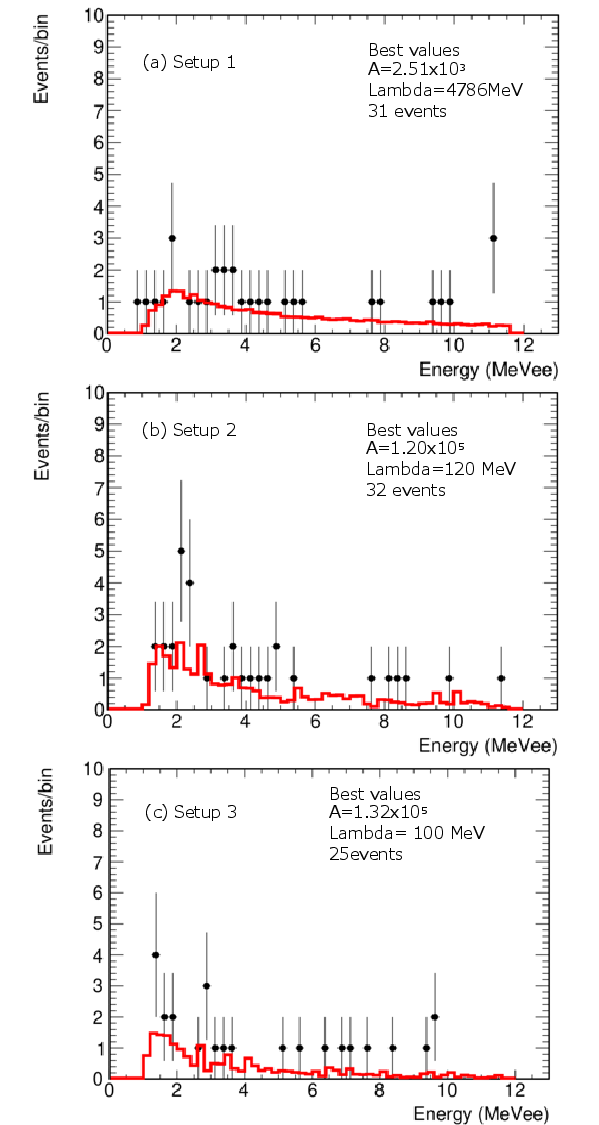}
    \caption{
    Observed deposited energy spectra in electron recoil equivalent energy for setups 1, 2, and 3 are shown in (a), (b), and (c), respectively. The dots represent the experimental data, and the red lines indicate the best-fit simulations.
    }
    \label{fig.observed_spectrum}
    \end{figure}

Figures~\ref{fig.observed_spectrum}~(a), (b), and (c) (bottom) show the observed energy distributions for the LiqSci detector along with the corresponding simulation spectra for setups 1, 2, and 3, respectively. The simulation spectra represent the best-fit results. The resulting \(\chi^2/{\rm dof}\) values for setups 1, 2, and 3 are 49.1/58, 48.1/58, and 46.9/58, respectively.

\begin{figure}[hbtp]
    \centering
    \includegraphics[width=\defaultimagesize\textwidth]{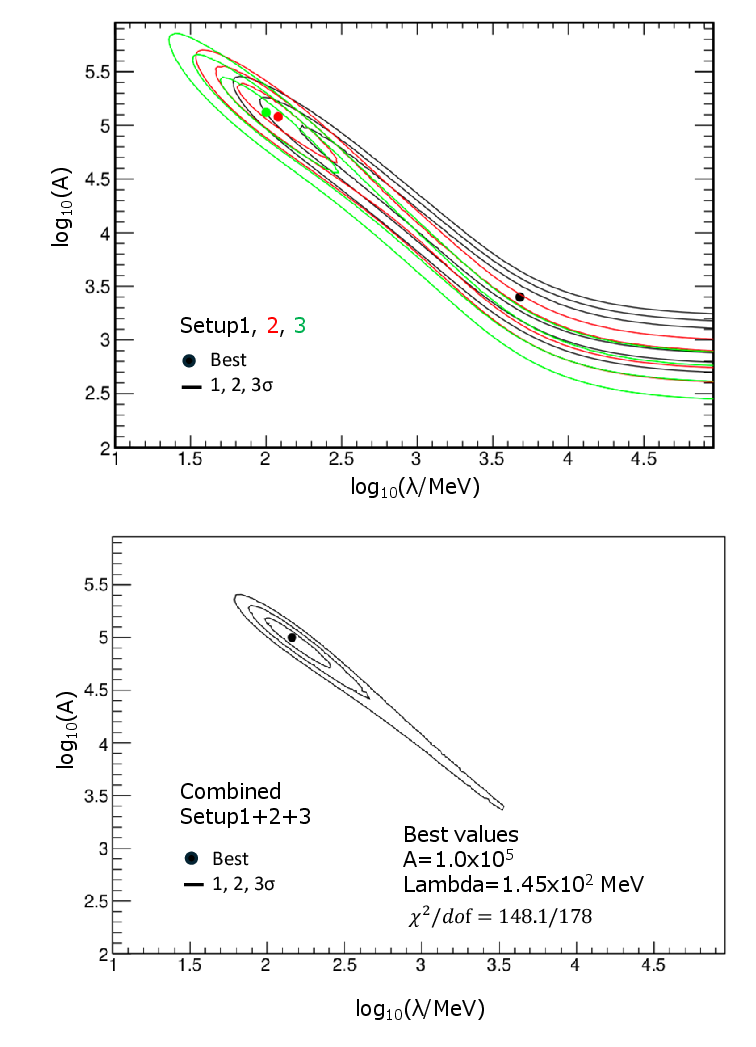}
    \caption{
    Distribution of \(\chi ^{2}\) for the fits to setups 1, 2, and 3 (top) and for the combined dataset (bottom).
    }
    \label{fig.chi2map}
    \end{figure}

Figure~\ref{fig.chi2map} (top) shows the \(\chi ^{2}\) distributions (or maps) for setups 1, 2, and 3. The parameters \(A\) and \(\lambda \) exhibit a strong correlation, and the overall trend of the \(\chi^{2}\) distribution is consistent across all setups. Therefore, the three datasets are combined by simply summing their \(\chi ^{2}\) values, \(\chi^2=\sum\chi^2_{\rm setup}\), as shown in Fig.~\ref{fig.chi2map} (bottom).The best-fit values are determined to be \(A=1.0\times10^5\) and \(\lambda=1.45\times10^2~{\rm MeV}\) with a total \(\chi^2/{\rm dof} = 148.1/178\).

    %
    %

\subsection{Neutron flux}
\label{sec.neutron_flux}

The neutron flux is estimated as \(\phi=A\lambda\), which corresponds to the integral of Eq.~(\ref{eq1}) over the neutron energy \(E_{n}\) from zero to infinity. The resulting fluxes for setups 1, 2, and 3 are determined to be \(\left(1.2_{-0.3}^{+10.5}\right)\), \(\left(1.44_{-0.31}^{+0.50}\right)\), and \(\left(1.32^{+0.38}_{-0.40}\right) \times 10^{-7} ~ {\rm cm^{-2}~ s^{-1} ~ POT^{-1}}\), respectively. These fluxes are consistent with each other, and the combined neutron flux is determined to be \(\left(1.45^{+0.22}_{-0.24}\right) \times 10^{-7} ~ {\rm cm^{-2}~ s^{-1} ~ POT^{-1}}\) with a total exposure of \(2.972\times10^{20}~\rm {POT}\).

   %
    %
    
\subsection{Systematic uncertainties}
\label{sec.sys.error}

We considered the systematic uncertainty sources as follows: the uncertainty in the quenching factor of the LiqSci detector, the energy resolution uncertainty from the source calibration, and the PSD distribution uncertainty from the beam-test calibration. The systematic uncertainty arising from the detector location is negligible compared to the above sources.
Systematic uncertainties were evaluated for each of the parameters implemented in the simulation---(a) the quenching factor curve as a function of proton recoil energy cited from literature, (b) the fitting function of the energy resolution obtained from calibration, and (c) the PSD distribution in the liquid scintillator---by varying each parameter within its $1\sigma$ uncertainty and determining the resulting variation in the converged neutron flux values.
As a result, the systematic uncertainties associated with the quenching factor, energy resolution, and PSD distribution were determined to be 9\%, 8\%, and 36\% relative to the central value of the neutron flux, respectively. The total systematic uncertainty was evaluated to be \(0.55\times10^{-7}~{\rm {cm}^{-2}~s^{-1}~{POT}^{-1}}\).

\subsection{Discussion}
\label{sec.discussion}

In this study, we collected data across three types of setups and varying beam powers (650~kW and 700--800~kW). All measured neutron fluxes were found to be consistent. These results indicate that the neutron flux per POT is invariant and independent of the beam power
within experimental uncertainties.
While the current data were taken in the "neutrino" mode, our next step is to determine the neutron flux in the "anti-neutrino" mode and investigate its dependence on the detector location within the experimental hall. By doing so, we aim to locate the position where the neutron background flux is minimized, which will guide the design of a future short-baseline NCQE experiment.

\section{Conclusion}
\label{sec.conclusion}

Neutrino-nucleus neutral-current quasielastic scattering (NCQE) events are masked by a large background of neutron-nucleus scattering events. These neutrons are produced in the surrounding material (such as sand and concrete) between the beam dump and the experimental area where the detectors are located. 
We performed measurements of these background neutron events using
a liquid scintillation detector,
equipped with a plastic scintillation detector as an active veto counter, at the J-PARC neutrino facility.
A dataset corresponding to \(2.972\times10^{20}~\rm{POT}\) was obtained in the neutrino mode. By selecting pulse-shape-discrimination (PSD) events in the liquid scintillation detector, we observed 88 proton-recoil events from neutron interactions with electron-equivalent energies of 0.98--11.60~MeV. By taking into account the detector efficiency and resolution and comparing the data with simulations, the neutron flux was determined to be \(\left[1.45^{+0.22}_{-0.24}~(\rm stat.)\pm 0.55~(\rm sys.)\right] \times10^{-7}~\rm cm^{-2}~s^{-1}~POT^{-1}\), assuming that the neutron energy spectrum follows an exponential function. This result will contribute to future short-baseline NCQE experiments.

\section*{Acknowledgments}

This work was supported by the KEK/J-PARC neutrino group
and the RCNP accelerator staffs in Osaka University, Japan.
This work was supported by 
the Neutron Measurement Consortium for Underground Physics,
JSPS KAKENHI Grant Grants, 
Grant-in-Aid for Scientific Research (C) [grant number 20K03998]
and
Transformative Research Areas (A) [grant number 24H02243]

\appendix
\renewcommand{\thefigure}{A.\arabic{figure}}
\setcounter{figure}{0}
\renewcommand{\thetable}{A.\arabic{table}}
\setcounter{table}{0}

\section*{Appendix A. Energy calibration of the Liquid scintillation detector}
\label{sec.appendixA}
\label{sec.LScalib}

\begin{figure}[hbtp]
    \centering
    \includegraphics[width=\defaultimagesize\textwidth]
    {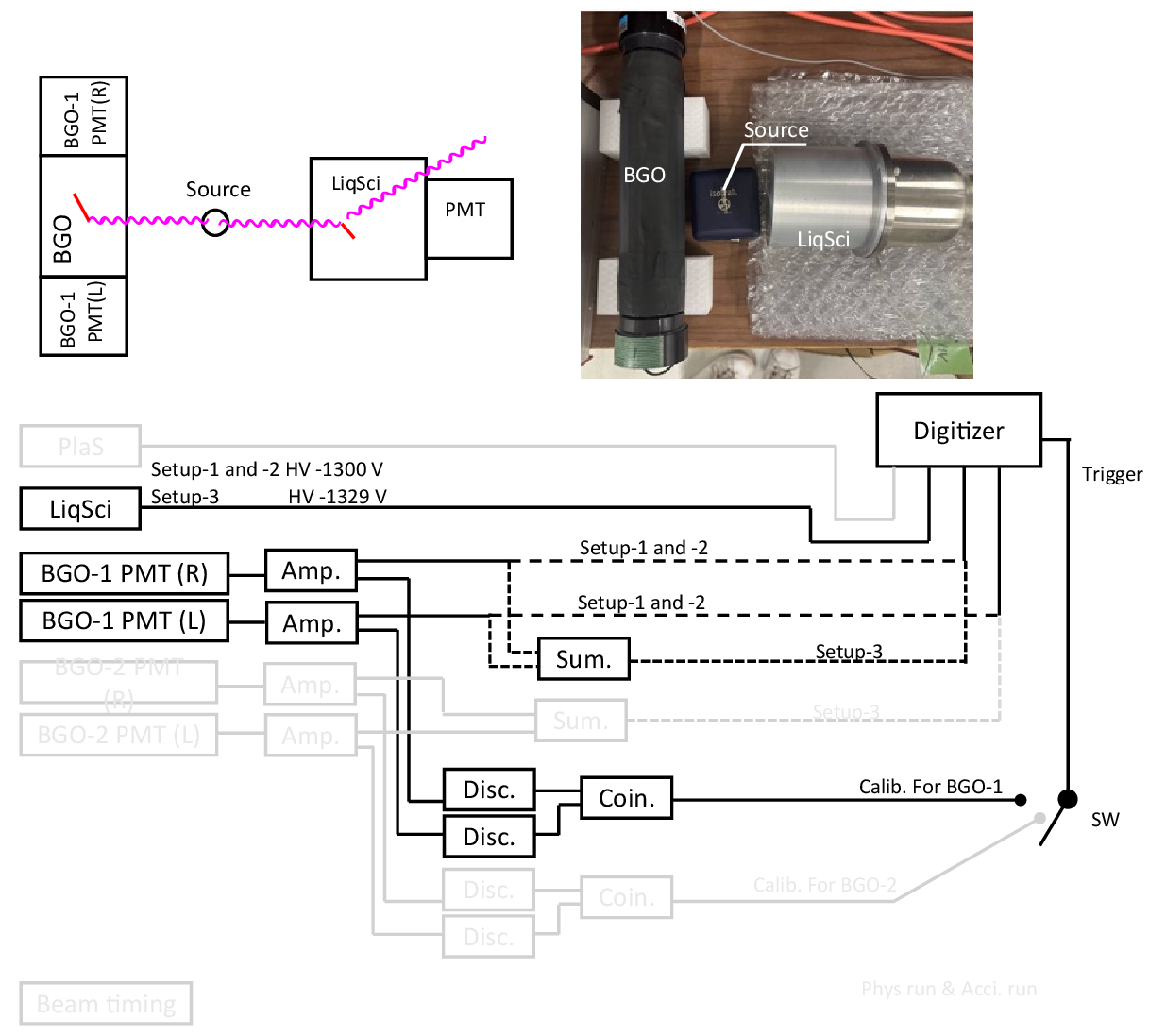}
    \caption{
    Setup for calibration of the Liquid scintillation detector.
    Top left is schematic view of location for source, BGO, and LiqSci detectors.
    Top right is graphics view.
    Bottom is the active data acquisition system.
    }
    \label{fig.setup.LS.calib}
    \end{figure}

As described in Section~\ref{sec.LiqSci}, we used \({}^{22}\)Na and \({}^{60}\)Co gamma-ray sources for the energy calibration of the LiqSci detector, with the experimental setup illustrated in Fig.~\ref{fig.setup.LS.calib}. These sources emit coincident gamma rays: two 511~keV gamma rays from \({}^{22}\)Na (via positron annihilation), and 1173 and 1332~keV gamma rays from \({}^{60}\)Co. Each source was positioned between the BGO and LiqSci detectors, and the data acquisition was triggered by the coincidence signal from the two PMTs of the BGO detector, allowing us to record the waveforms from both detectors.

    %

By tagging photoelectric absorption events corresponding to 511, 1173, or 1332~keV in the BGO detector, events where the counterpart gamma ray enters the LiqSci detector are selected. This selection enables a direct comparison between the experimental data and a simplified simulation. In the simulation, mono-energetic gamma rays were generated to enter the LiqSci geometry, and the deposited energy was recorded on an event-by-event basis. The simulated spectra were then fitted to the data by varying the energy-to-charge conversion coefficient \(k\) and the energy resolution, the latter parameterized as \(\sigma = \sigma_1 E^{1/2} + \sigma_0\), where \(E\) is the deposited energy in units of MeV electron equivalent.

\begin{figure}[hbtp]
    \centering
    \includegraphics[width=\defaultimagesize\textwidth]
    {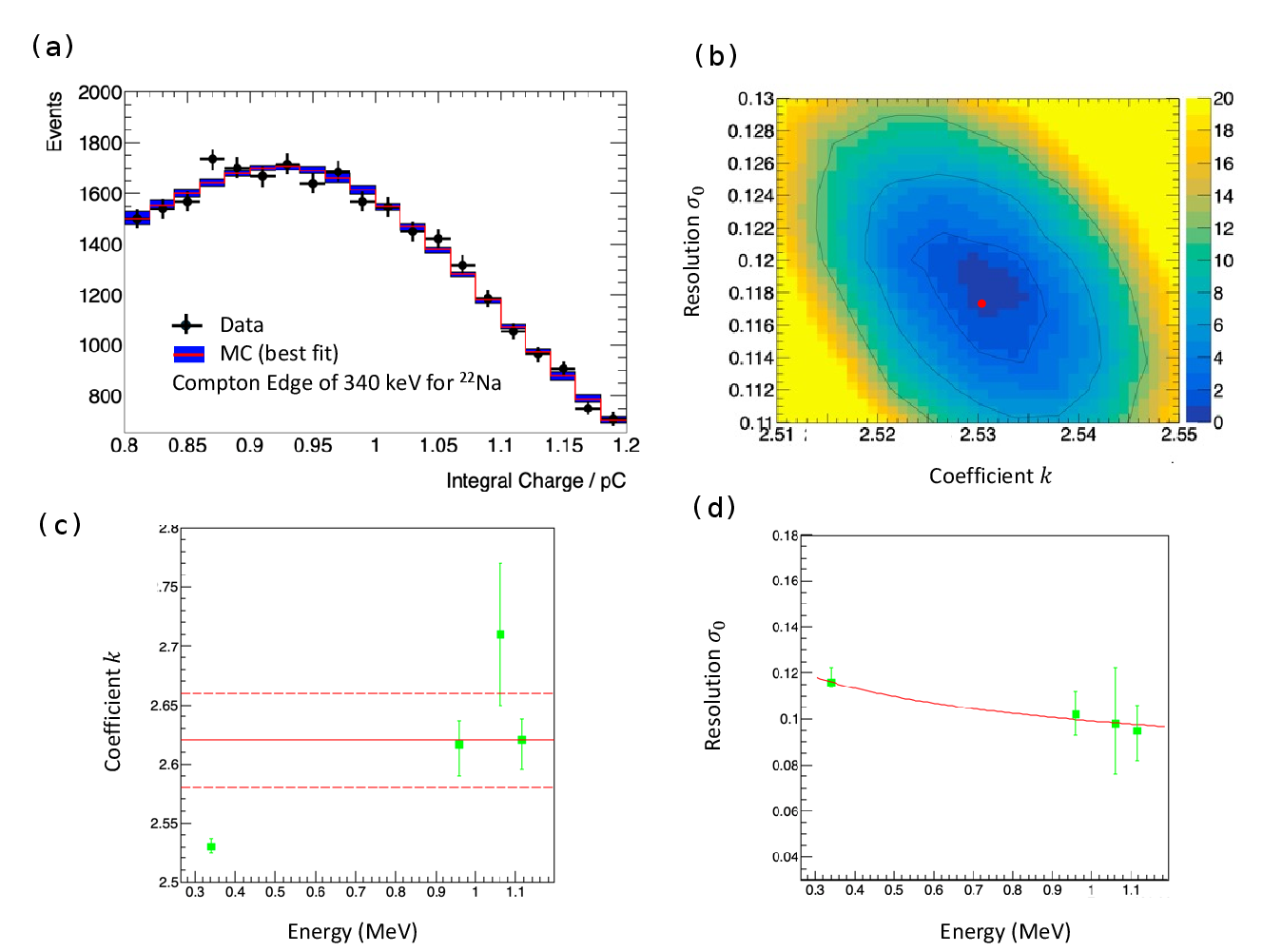}
    \caption{
    (a) Measured charge distribution for Compton scattering (~341~keV edge) using a $^{22}$Na source, compared with the best-fit simulation (MC) for setups 1 and 2. (b) $\chi^2$ distribution as a function of the parameters $k$ and $\sigma$. (c) Best-fit values of $k$ versus energy. The red solid and dashed lines represent the error-weighted average and its associated uncertainty interval, respectively.(d) Best-fit values of $\sigma$ versus energy. The red curve indicates the fit to the data using the function \(\sigma=\sigma_1 E^{1/2}+\sigma_0\).
    }
    \label{fig.result.LS.calib}
    \end{figure}

Figure~\ref{fig.result.LS.calib}~(a) shows the measured charge distribution for the Compton scattering events, featuring a Compton edge at \(\sim \)340~keV, compared with the simulated spectrum. Figure~\ref{fig.result.LS.calib}~(b) shows the \(\chi ^{2}\) distribution as a function of the parameters \(k\) and \(\sigma \). For the Compton edges at 963 and 1118~keV obtained from the \({}^{60}\)Co source, the best-fit values were determined using the same methodology applied to the \(\sim \)340~keV edge. Figures~\ref{fig.result.LS.calib}~(c) and (d) display the calibrated values of \(k\) and \(\sigma \) as a function of energy \(E\) for setups 1 and 2, respectively. The optimal parameters for \(\sigma _{1}\) and \(\sigma _{0}\) were then extracted by fitting the parameterized function to the \(\sigma \) versus \(E\) plot. Setup 3 was analyzed using the same procedure. The resulting best-fit values of \(k\), \(\sigma _{0}\), and \(\sigma _{1}\) for setups 1, 2, and 3 are summarized in Table~\ref{table.calib.LS}. These calibration factors and their associated uncertainties were subsequently implemented into the detector response model within the simulation.

    %

    %
    %
    \begin{table}[htbp]
    \centering
    \caption{Summary of the best values $k$, $\sigma_0$, and $\sigma_1$ for LiqSci detector calibration with high voltage supply of $-$1300 and $-$1329~V.}
    \begin{tabular}{c c c }
    \hline
    \hline
    &$V=-1300~\rm V$ & $V=-1329~\rm V$ \\
    & (setup 1 and 2) &  (setup 3) \\
    \hline
    $k ~[\rm pC/MeV]$ & $2.62 \pm 0.04$ & $3.11 \pm 0.04$\\
    $\sigma_1 ~{[\rm {MeV}^{1/2}]}$ & $(9.6 \pm 0.8)\times10^{-2}$ & $(8.3 \pm 0.7)\times10^{-2}$ \\
    $\sigma_0$ & $-(1.7 \pm 0.9)\times10^{-1}$ & $-(2.2 \pm 1.0)\times10^{-1}$ \\
    \hline
    \hline
    \end{tabular}
    \label{table.calib.LS}
    \end{table}

\appendix
\renewcommand{\thefigure}{B.\arabic{figure}}
\setcounter{figure}{0}
\section*{Appendix B. Liquid scintillation detector PSD calibration in beamtest}
\label{sec.RCNP}

We performed a calibration of the proton-recoil event selection criteria for pulse-shape discrimination (PSD) at the Research Center for Nuclear Physics (RCNP), Osaka University, Japan. Quasi-monoenergetic 65~MeV neutrons were produced via the \(\rm {^{7}Li}(p,n){^{7}Be}\) reaction, using a 10-mm-thick lithium target and a proton beam operated at a maximum current of \(\rm 1~\mu A\).The resulting neutron energy spread was 2.2~MeV (FWHM). The data acquisition (DAQ) system utilized a simple self-triggering scheme for the liquid scintillation detector (LiqSci). Two different high voltages, \(-1300~\rm V\) and \(-1329~\rm V\), 
were applied to the PMT of LiqSci detector.

    %
    %
    \begin{figure}[hbtp]
    \centering
    \includegraphics[width=\defaultimagesize\textwidth]
    {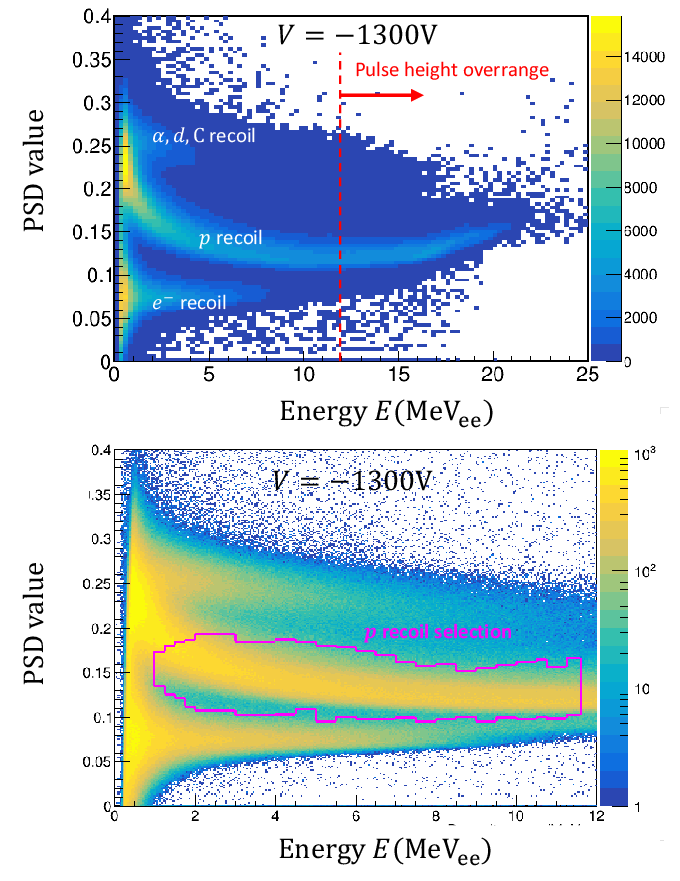}
    \caption{   
    Top panel: Correlation between the deposited energy and PSD value at a high voltage of \(-1300\)~V for the LiqSci detector, where three distinct components are observed. 
    Above an energy of 12~\(\mathrm{M}eV_{ee}\), the pulse height exceeds the dynamic range (or saturates).
    The energy corresponds to $11.60~\rm MeV_{ee}$ for the run at a high voltage of $-1329$~V.
    Bottom panel: Selection criteria for proton-recoil events overlaid on the PSD distribution.
    }
    \label{fig.PSD.RCNP}
    \end{figure}

Figure~\ref{fig.PSD.RCNP} shows the correlation between the deposited energy and the PSD value at an applied voltage of $-1300~\rm V$ for the LiqSci detector. The definition of PSD is described in Sec.~\ref{sectoin.PSD}, and its calculation details can be found in Refs.~\cite{Ommura_2023, ITO2023168701}. We identified three distinct components corresponding to electron recoil, proton recoil, and heavier-particle recoil. Due to the pulse height dynamic range of the digitizer, signal saturation occurs above 12~\(\mathrm{MeV}_{\rm ee}\). The electron-recoil component predominantly arises from gamma-ray interactions, whereas the other recoil components originate from neutron interactions. However, the quenching factors for heavier particles (such as carbon nuclei, alpha particles, and deuterons) suffer from large uncertainties. Therefore, we adopted an approach that compares only the proton-recoil events with the simulation.

\begin{figure*}[hbtp]
    \centering
    \includegraphics[width=0.9\textwidth]{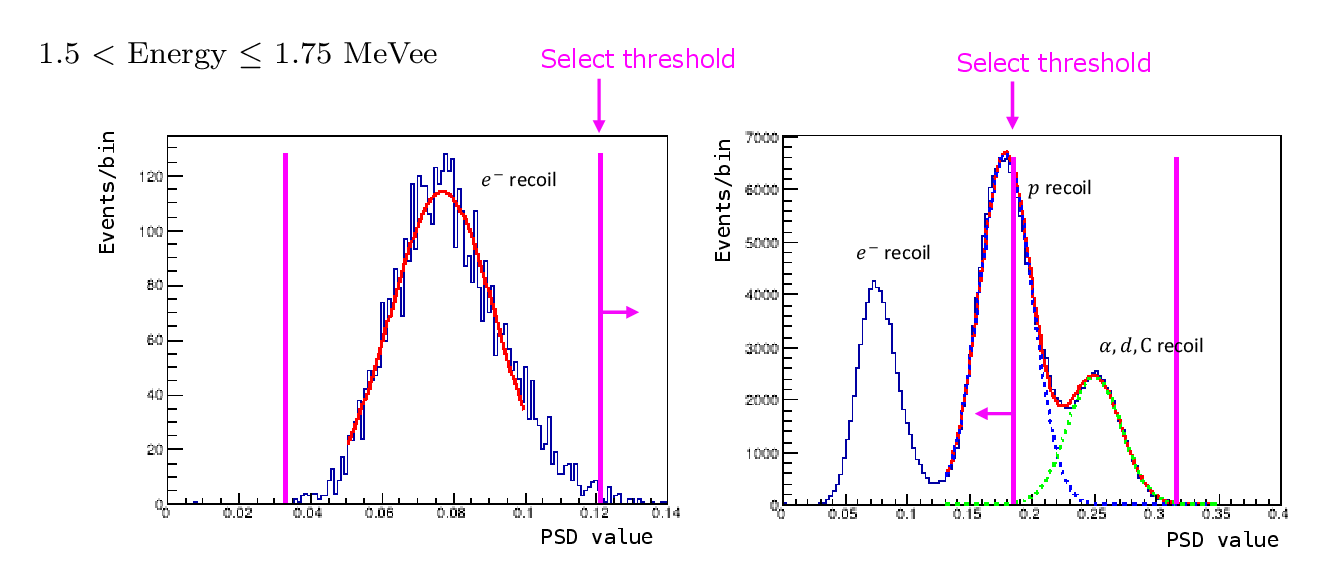}
    \caption{
    Examples of PSD distributions ($1.5\text{--}1.75~\text{MeV}_{\text{ee}}$) with Gaussian fits for threshold determination, showing (left) the $\gamma$-ray calibration setup to eliminate electron recoils at the $3\sigma$ level and (right) the neutron beam setup to eliminate heavy-nucleus recoils at the $3\sigma$ level.
    }
    \label{fig.PSD.cut.RCNP}
    \end{figure*}

To determine the PSD selection criteria for proton-recoil events, the two-dimensional distribution was projected onto a one-dimensional PSD histogram for each energy interval. 
The thresholds for excluding electron-recoil and heavier-particle events were determined at the \(3\sigma\) level by fitting Gaussian functions to the respective components
(see Fig.~\ref{fig.PSD.cut.RCNP}, e.g., $1.5\text{--}1.75~\text{MeV}_{\text{ee}}$).
The same methodology was applied to the data taken at \(V=-1329~\rm V\). The gain shift resulting from this higher voltage caused a shift in the saturation boundary to 11.60~\(\mathrm{MeV_{ee}}\). To ensure consistent analysis across all configurations, a common energy range below 11.60~\(\mathrm{MeV_{ee}}\) was adopted for setups 1, 2, and 3.

    %
    %

\def\urlprefix{}
\bibliography{Reference.bib}

\end{document}